\documentclass[12pt,epsfig]{article}
\usepackage[latin1]{inputenc}\usepackage{epsfig}\usepackage[english]{babel}
\newcommand{\doublespace}{\renewcommand{\baselinestretch}{1.75}
\Large\normalsize}
\doublespace
\def\flex{\raise 8pt\hbox{$\leftrightarrow $}\! \! \! \! \! \! }

\begin{document}
\begin{center}
{\large\bf Support of dS/CFT correspondence from space-time perturbations}
\end{center}
\vspace{1ex}
\centerline{\large
E. Abdalla$^{\ast (1)}$, K. H. C. Castello-Branco$^{\ast (2)}$ and
A. Lima-Santos$^{\dag (3)}$}
\begin{center}
$^{\ast}$Universidade de S\~ao Paulo, Instituto de F\'\i sica\\
Caixa Postal 66318, 05315-970, S\~ao Paulo-SP, Brazil;\\
$^{\dag}$Universidade Federal de S\~ao Carlos, Departamento de
F\'{\i}sica \\
Caixa Postal 676, 13569-905, S\~ao Carlos-SP, Brazil\\
\end{center}
\vspace{6ex}

\begin{abstract}

We analyse the spectrum of perturbations of the de Sitter space on
the one hand, while on the other hand we compute the location of the poles
in the Conformal Field Theory (CFT) propagator at the border. The coincidence 
is striking, supporting a dS/CFT correspondence. We show that the
spectrum of thermal excitations of the CFT at the past boundary
$I^{-}$ together with that spectrum at the future
boundary $I^{+}$ is contained in the
quasi-normal mode spectrum of the de Sitter space in the bulk.
\end{abstract}
\vspace{4ex}
(1)eabdalla@fma.if.usp.br\\
(2)karlucio@fma.if.usp.br\\ 
(3)dals@df.ufscar.br

\newpage
\section{Introduction}
The main goal of string theory nowadays is to prove itself capable
of coping with experimental evidences. On the other hand, recently, it has
been made almost evident from supernova observations \cite{lbpos} as well 
as from the data supplied by the COBE satelite \cite{cobe}, that we live 
in a flat world with a kind of special matter, or  possibly a positive 
cosmological constant, {\it i.e.}, de Sitter space-time \cite{bousso1}.
It is thus annoying that several results in string theory
are obtained only in Anti-de Sitter (AdS) space, namely a space with
a negative cosmological constant, especially the celebrated correspondence
between the bulk AdS space and the Conformal Field Theory (CFT) at the 
border \cite{malda,severalads}. It is even deceiving, in view of
the above observations, that de Sitter space poses serious questions for
the formulation of string theory \cite{witten}. It is our aim
here to contribute to the extension of the bulk to border correspondence to the
case where the bulk cosmological constant is positive, that is, 
to investigate whether the de Sitter space in the bulk has any
relationship with a purported CFT at its border. 

In fact, a holographic duality relating quantum gravity on 
$D$-dimensional de Sitter space ($dS_D$) to a CFT residing on the
past boundary ($I^{-}$) of $dS_{D}$ has recently been proposed by 
Strominger in \cite{Strom1}, what has motivated several works
discussing the setup of this duality (see, for instance, 
\cite{severalds}). Moreover, as shown by Gibbons and Hawking 
\cite{GibHawk} the de Sitter space cosmological event horizon is endowed 
with entropy and temperature, in analogy with a black hole event 
horizon. Thus, quantum gravity, as a natural description of black
holes, has to be extended to the whole universe \cite{bousso1}.

We have already analysed the question raised above in the case
of three-dimensional de Sitter space, where both, perturbations of $dS_{3}$
solution in the bulk, as well as the two-dimensional CFT propagator
spectrum at the border, are explicitly computable \cite{ablimbin}. There
we obtained both spectra, and they turned out to coincide exactly. Indeed,
the solution of the Klein-Gordon equation in the bulk of the de Sitter
space leads to a hypergeometric equation, which, upon employing the
boundary condition imposing the vanishing of the field at the event
horizon, leads to a four-fould set of quasi-normal modes. Inspection of 
the propagator of the CFT at the boundary space, for the physically 
realizable boundary conditions of the fields, shows a spectrum that exactly
coincides with the previously described one. 

Nevertheless, three-dimensional de Sitter space-time is far simpler than real
systems and the actual problem remains basically
unsolved. Furthermore, three-dimensional de Sitter space does not admit
a black hole event horizon (there is no black hole solution), but only 
a cosmological horizon. On the other hand,  four-dimensional
Schwarzschild-de Sitter solution displays a very peculiar spectrum
of perturbations. There are quasi-normal modes at short times,
followed by a power law decay typical of the asymptotically flat 
Schwarzschild solution, and finally by an exponential tail 
\cite{brady,abdmolsaa}. Our suspicion is that only this exponential tail is 
reflected in the CFT at the boundary. For the moment the problem is too 
complex, and we remain for the time being on perturbations of the empty 
dS space, which is exactly computable, comparing with the corresponding 
CFT results at the border. The conclusions should apply for small black
holes, in view of the above peculiar spectrum of perturbations.

In summary, a qualitative correspondence between quasi-normal modes in
AdS spaces and the decay of perturbations in the dual CFT has been 
obtained \cite{ref5deablimbin,ref6deablimbin,ref7-8deablimbin,ref9deablimbin,
ref10deablimbin}. In the case of an AdS space-time as described
by the BTZ black hole \cite{btz} a precise agreement between
quasi-normal mode frequencies and the location of the poles in the
retarded correlation function of the corresponding perturbations in the
dual CFT has been obtained in \cite{birmg}, whereas in \cite{ablimbin} 
a similar correspondence was found for three-dimensional dS space. The 
investigation of the validity of this agreement for four-dimensional dS 
space, as well as its extension for higher dimensions, is thus an 
important question. Therefore, our aim is to generalize the previously 
obtained three-dimensional results, as well as reconfirming the many
indications of such a holographic duality for de Sitter space.

\section{The Geometry of the de Sitter Space-Time}
The $D$-dimensional de Sitter space-time, $dS_{D}$, can be visualized as the
hyperboloid
\begin{equation}
\eta_{AB}X^{A}X^{B}\equiv -(X^{0})^{2} +(X^{1})^{2} + ... +
(X^{D})^{2} = a^{2}\quad ,           \label{hiperb}
\end{equation}
embedded in $(D+1)$-dimensional Minkowski space-time with metric
\begin{equation}
ds^{2} = \eta_{AB}dX^{A}dX^{B}\quad ,            \label{metr}
\end{equation}
where $\eta_{AB} = diag(-1,1,...,1)$ and $A,B = 0,1,...,D$. The
parameter $a$ is the de Sitter radius.

The $dS_{D}$ metric is induced from the flat
metric (\ref{metr}) on the hypersurface (\ref{hiperb}). It is a solution
of the Einstein field equations with a positive cosmological constant 
$\Lambda$, related to $a$ by  $\Lambda = (D-2)(D-1)/2a^2$ \cite{Leshouches}.

Several coordinate systems used to describe the structure of 
$dS_{D}$ are discussed in \cite{Leshouches}. Among
these, the static coordinates are particularly useful, since they
make the existence of a (cosmological) event horizon manifest
in space-time, being more suitable for the purpose of this work.

As in \cite{Klem1}, we fix
\begin{equation}
(X^{0})^{2} - (X^{D})^{2} = - a^{2}V(r)\quad ,         \label{param}
\end{equation}
where
\begin{equation}
V(r) = 1 - \frac{r^{2}}{a^{2}}\quad .     \label{funcmetr}
\end{equation}

Then, we use (\ref{hiperb}) to arrive at the constraint  
\begin{equation}
(X^{1})^{2} + ... + (X^{D-1})^2 = r^{2}\quad ,         \label{somaquadr}    
\end{equation}
so that the coordinates $X^{1},...,X^{D-1}$ range over a
$(D-2)-$sphere, $S^{D-2}$, of radius $r$. By parametrizing the
hyperbola (\ref{param}) by 
\begin{equation}
X^{0}= \sqrt{r^{2}-a^{2}}\cosh \frac{t}{a}\quad , \label{param1}  
\end{equation}
\begin{equation}
X^{D}= \sqrt{r^{2}-a^{2}}\sinh \frac{t}{a}\quad ,       \label{param2}
\end{equation}
the induced metric on the hypersurface (\ref{hiperb}) is given by
\begin{equation}
ds^{2} = -V(r)dt^{2} + V^{-1}(r)dr^{2} + r^2d\Omega^{2}_{D-2}\quad ,
\label{metr2}
\end{equation}
where $d\Omega^{2}_{D-2}$ is the metric of the unit sphere 
$S^{D-2}$. This is the form of the $dS_{D}$ metric in static
coordinates. From (\ref{metr2}) it is easy to see that an observable
at $r=0$ is surrounded by an event horizon at $r=a$.
The static coordinates do not cover the whole space-time, but only
the interior region of the cosmological horizon, which corresponds 
to the left triangle in the Penrose-Carter diagram shown below. The
past ($I^{-}$) and future ($I^{+}$) event horizons correspond to 
$r=\infty$ \cite{GibHawk}.

\begin{figure}[hbt]
\begin{center}
\mbox{\epsfig{file=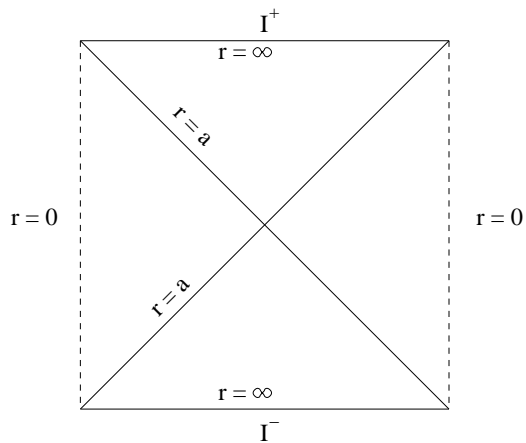,width=0.5\textwidth}}
\end{center}
\caption{The Penrose-Carter diagram of de Sitter space-time.}
\label{fig1}
\end{figure}

In view of the computation of the de Sitter invariant Hadamard two-point 
function it is convenient to define an invariant $P(X,X^{\prime})$ 
associated to two points $X$ and $X^{\prime}$ in de Sitter space. We define
\begin{equation}
a^{2}P(X,X^{\prime})=\eta_{AB}X^{A}X^{\prime B}\quad ,\label{defP1}
\end{equation}
which is related to the geodesic distance $d(X,X^{\prime})$ between
two points by $P=\cos(d/a)$. In static coordinates, we can easily
express $P(X,X^{\prime})$ as 
\begin{equation} 
a^{2}P(X,X^{\prime})= -\sqrt{r^{2}-a^{2}}\sqrt{r^{\prime 2}-a^{2}}\cosh \frac
{t-t^{\prime}}{a} + rr^{\prime}\cos \Theta\quad ,              
\label{defP2}
\end{equation} 
where $\Theta= \Theta(\Omega,\Omega^{\prime})$ denotes the geodesic
distance between two points on the unit sphere $S^{D-2}$. Later we will
restrict our results to the case $D=4\,$, where
\begin{equation}
\cos \Theta = \sin \theta \sin \theta^{\prime} \cos (\phi - \phi^{\prime})
+ \cos \theta \cos \theta^{\prime}\quad    \label{ang}
\end{equation} 
holds, as can easily be obtained in terms of the usual polar and azimuthal
angles $\theta$ and  $\phi$.

\section{Scalar Perturbations of the  $dS_{D}$ Space}
Scalar perturbations of the $dS_{D}$ space-time are described
by the Klein-Gordon wave equation
\begin{equation}
\frac{1}{\sqrt{-g}}\partial _{\mu }\left( \sqrt{-g}g^{\mu \nu }\partial
_{\nu }\Phi \right) -\mu ^{2}\Phi=0\quad ,             \label{K-G} 
\end{equation}
where $\mu$ is the mass of the perturbing scalar field $\Phi$
and $g_{\alpha \beta}$ is the $dS_{D}$ metric,
\begin{equation}
g_{\mu \nu }=diagonal(-V(r),V^{-1}(r),r^{2}\hat{1}_{D-2})\quad ,   
\label{diag}
\end{equation}
where $\hat{1}_{D-2}$ is the identity matrix on $S^{D-2}$. Therefore,
$\sqrt{-g}=\sqrt{-\det(g_{\alpha \beta})}=
r^{(D-2)}\sin^{D-3}\theta_{1},\sin^{D-4}\theta_{2}...\sin\theta_{D-3}$
and from (\ref{K-G}) we have
\begin{eqnarray}
&&\frac{1}{r^{D-2}}\partial _{t}\left( r^{D-2}(-V^{-1}(r))\partial _{t}\Phi
(t,r,\Omega) \right) +\frac{1}{r^{D-2}}\partial _{r}\left(
r^{D-2}V(r)\partial _{r}\Phi (t,r,\Omega)\right) + \nonumber \\
&& + \frac{1}{r^{2}}\bigtriangledown _{_{D-2}}^{2}\Phi (t,r,\Omega
) - \mu ^{2}\Phi(t,r,\Omega) = 0\quad ,                   \label{sep1} 
\end{eqnarray}
where $\Omega$ denotes the set of angular variables, and 
$\bigtriangledown _{_{D-2}}^{2}$ corresponds to the angular
derivatives, that is, the Laplacian on the unit sphere $S^{D-2}$.

The above equation is separable, with the Ansatz 
\begin{equation}
\Phi (t,r,\Omega )=e^{-i\omega t}R_{\ell}(r)Y^{m}_{\ell}(\Omega)\,,
\label{sep2}
\end{equation}
where $Y^{m}_{\ell}(\Omega)$ are the (hyper-)spherical harmonics on $S^{D-2}$,
which obey 
\begin{equation}
\bigtriangledown _{_{D-2}}^{2}Y^{m}_{\ell}(\Omega)=
-\ell(\ell+D-3)Y^{m}_{\ell}(\Omega)\,.                    \label{Lapl}
\end{equation}
In $Y^{m}_{\ell}$, ${\ell}$ is a positive integer and ${m}$ is a collective
index ($m_{1},m_{2},...,m_{D-3}$) \cite{bateman}.

Therefore, we have for the radial dependence in $R_{\ell}\,$,
\begin{equation}
\frac{V(r)}{r^{D-2}}\frac{d}{dr}\left( V(r)r^{D-2}\frac{dR(r)}{dr}\right) +%
\left[ \omega ^{2}-V(r)\left( \frac{\ell(\ell+D-3)}{r^{2}}+\mu ^{2}\right) \right]
R(r)=0,                 \label{deprad}
\end{equation}
where we have omitted the index $\ell$ in $R(r)$.

Defining
\begin{equation}
z=\frac{r^{2}}{a^{2}}\,,                    \label{novar}
\end{equation}
equation (\ref{deprad}) can be written as 
\begin{eqnarray}
&&\frac{4}{a^{2}}z(z-1)^{2}\frac{d^{2}R(z)}{dz^{2}}+\frac{2}{a^{2}}%
(z-1)[z(D+1)-(D-1)]\frac{dR(z)}{dz}+ \nonumber \\
&& + \left[ \omega ^{2}-(1-z)\left( \frac{%
l(l+D-3)}{a^{2}z}+\mu ^{2}\right) \right] R(z)=0\,.           \label{depz2}
\end{eqnarray}
Now, by setting the Ansatz
\begin{equation}
R=z^{\alpha }(1-z)^{\beta }F(z)\quad,       \label{ansatz}
\end{equation}
equation (\ref{depz2}) can be transformed into
\begin{eqnarray}
&&z(z-1)\frac{d^{2}F(z)}{dz^{2}}+\frac{1}{2}\left[ (4\alpha +4\beta
+D+1)z-(4\alpha +D-1)\right] \frac{dF(z)}{dz} + \nonumber \\
&&+\left( (\alpha +\beta )^{2}+\frac{D-1}{2}(\alpha +\beta )+\frac{\mu
^{2}a^{2}}{4}\right) F(z) + \nonumber \\
&&+\frac{(4\beta ^{2}+\omega ^{2}a^{2})z-(4\alpha ^{2}+2\alpha
(D-3)-\ell(\ell+D-3))(z-1)}{4z(z-1)}F(z) = 0\,,\nonumber \\               
\label{eqtransf}
\end{eqnarray}
which is a hypergeometric equation, that is,
\begin{equation}
z(z-1)\frac{dF(z)}{dz}+\left[(1+A+B)z-C\right]
\frac{dF(z)}{dz}+ABF(z)=0 \,,              \label{hyperg}
\end{equation}
provided we choose the arbitrary constants $\alpha$ and $\beta$ as
\begin{eqnarray*}
4\alpha ^{2}+2\alpha (D-3)-l(l+D-3) &=&0 \,, \\
4\beta ^{2}+\omega ^{2}a^{2} &=&0 \,,                \label{eqalfabeta}
\end{eqnarray*}
whose solutions are
\begin{eqnarray*}
{\rm (i)} &:&\qquad \alpha =\frac{l}{2}\,,\qquad \beta =\frac{ia\omega }{2}\quad;\\
{\rm (ii)} &:&\qquad \alpha =\frac{l}{2}\,,\qquad \beta =-\frac{ia\omega }{2}\quad;
\\
{\rm (iii)} &:&\qquad \alpha =-\frac{l+D-3}{2}\,,\qquad \beta =\frac{ia\omega 
}{2}\quad; \\
{\rm (iv)} &:&\qquad \alpha =-\frac{l+D-3}{2}\,,\qquad \beta =-\frac{ia\omega 
}{2}\quad.                     \label{casos}
\end{eqnarray*}

For case (i), the constants $A$, $B$, and $C$ are given by the set
\begin{equation}
A_{\mp}= B_{\pm}= 
\frac{l+ia\omega}{2} +\frac{1}{4}\left[D-1\pm
\sqrt{(D-1)^{2}-4\mu^{2}a^{2}}\right],
\label{valorAB}
\end{equation}
\begin{equation}
C= \ell+ \frac{D-1}{2}\,.                 \label{valorC}
\end{equation}

Therefore we have the two solutions for $R(z)\,$, that is 
\begin{equation}
R(z)=z^{l/2}(1-z)^{ia\omega /2}F(\frac{l+ia\omega +h_{-}}{2},\frac{%
l+ia\omega +h_{+}}{2},l+\frac{D-1}{2},z)             \label{solhyper1}
\end{equation}
and
\begin{equation}
R(z)=z^{l/2}(1-z)^{ia\omega /2}F(\frac{l+ia\omega +h_{+}}{2},\frac{%
l+ia\omega +h_{-}}{2},l+\frac{D-1}{2},z),            \label{solhyper2}
\end{equation}
where $F$ is the usual hypergeometric function \cite{gradst} and
\begin{equation}
h_{\pm }=\frac{1}{2}\left[D-1\pm \sqrt{(D-1)^{2}-4\mu ^{2}a^{2}}\right]\,.  
\label{parh} 
\end{equation}

The corresponding quasi-normal modes are solutions of the equations
\begin{equation}
C-A=-n\quad, \qquad C-B=-n\quad,                \label{eqmodos}
\end{equation}
{\it i.e.},
\begin{eqnarray*}
l+\frac{D-1}{2}-\frac{l+ia\omega +h_{\mp }}{2}+n &=&0\quad, \\
l+\frac{D-1}{2}-\frac{l+ia\omega +h_{\pm }}{2}+n &=&0\quad .   \label{sistema} 
\end{eqnarray*}

The corresponding quasi-normal frequencies are then given by
\begin{eqnarray}
ia\omega &=&2n+l+D-1-h_{\pm } \\
&=&2n+l+h_{\mp }\quad .                   \label{eqome1}
\end{eqnarray}
This means that we have the two sets of solutions
\begin{eqnarray}
\pm \,ia\omega _{R} &=&2n+l+h_{+}\quad,\qquad \pm \,ia\omega_{L}=2n+l+h_{-}\quad,   
\label{eqome2}
\end{eqnarray}
where we have included the complex conjugated solutions.

The case (ii) is obtained from the case (i) by complex conjugation and the
quasi-normal frequencies are solutions implying case (ii) as a
solution. Further quasi-normal modes are obtained from the symmetry
$\ell \rightarrow -(\ell + D - 3)$ , which implies cases (iii) and
(iv) above, or the set
\begin{eqnarray}
\pm\,ia\omega _{R}= 2n-(l+D-3)+h_{+}\,,\,
\pm\,ia\omega_{L}= 2n-(l+D-3)+h_{-}\quad.                  
\label{eqome3}
\end{eqnarray}

\section{Two-Point Correlator for CFT Operators at the Boundary}
We define a de Sitter 
invariant Hadamard two-point function as \cite{Strom1,Klem1}
\begin{equation}
G(X,X^{\prime})=const<0|{\Phi(X),\Phi(X^{\prime})}|0>\quad,     \label{green}
\end{equation}
which obeys
\begin{equation}
(\bigtriangledown _{_{X}}^{2} - \mu ^{2})G(X,X^{\prime}) = 0\quad,    
\label{eqgreen}
\end{equation}
where $\bigtriangledown _{_{X}}^{2}$ is the Laplacian on $dS_{D}$. The
Green function $G(X,X^{\prime})$ depends on $X$ and $X^{\prime}$ only
through the invariant $P(X,X^{\prime})$, the distance between the points
points $X$ and $X^{\prime}$, given by eq.(\ref{defP1}). We can write 
$G(X,X^{\prime}) = G(P(X,X^{\prime}))$, and from (\ref{eqgreen}) we 
obtain \cite{Can,Klem1}
\begin{equation}
(1-P^{2})\frac{d^{2}G}{dP^{2}} -DP\frac{dG}{dz} - \mu^{2}a^{2}G(P)=0\quad,
\label{eqGr}
\end{equation}
which, by means of the change of variable $z = (1+P)/2$ becomes a 
hypergeometric equation
\begin{equation}
z(1-z)\frac{d^{2}G}{dz^{2}}+ \left (\frac{D}{2} - Dz \right )\\
\frac{dG}{dz} -\mu^{2}a^{2}G(z)=0\quad.                 \label{eqhyper}
\end{equation}
The solution is the hypergeometric function $F$,
\begin{equation}
G(z) = Re F \left (h_{+}, h_{-},\frac{D}{2}; z \right)\quad,   
\label{real1}
\end{equation}
{\it i.e.}, 
\begin{equation}
G(P) = ReF \left (h_{+},h_{-},\frac{D}{2};\,\frac{1+P}{2}\right)\quad,   
\label{real2}
\end{equation}
where $h_{\pm}$ is given by (\ref{parh}).

In order to explicitly calculate the two-point correlator for an
operator coupled to the bulk field $\Phi$, we restrict now our
considerations to the four-dimensional case ($D=4$). The results can 
be naturally extended for any dimension. The two-point
correlator can be obtained analogously to \cite{Strom1,Klem1} from
\begin{eqnarray}
&&\lim_{r\rightarrow \infty}\int_{I^{-}}dt d\theta d\phi dt^{\prime}
d\theta^{\prime}d\phi^{\prime}\sin\theta\sin
\theta^{\prime}\frac{(rr^{\prime})^{2}}{a^{2}}\times \nonumber \\ 
&&{\left[\Phi(t,r,\theta,\phi)\flex\partial_{r_{\ast}}G(t,r,\theta,\phi;t^{\prime},r^{\prime},
\theta^{\prime},\phi^{\prime})\flex\partial_{r^{\prime}_{\ast}}
\Phi(t^{\prime},r^{\prime},\theta^{\prime},\phi^{\prime})\right]}_{r=r^{\prime}}\,,
\label{lim1}
\end{eqnarray}
where $dr_{\ast}=(-V(r))^{-1/2}dr$.

The asymptotic behaviour of $G(X,X^{\prime})$ at the conformal
boundary, in static coordinates, is given by (see equation (B.9) of 
\cite{Klem1})
\begin{equation}
\lim_{r,r^{\prime}\rightarrow
\infty}G(t,r,\theta,\phi;t^{\prime},r^{\prime},
\theta^{\prime},\phi^{\prime}) = c_{+}(rr^{\prime})^{-h_{+}}[\cosh 
(\frac{t-t^{\prime}}{a}) - \cos \Theta]^{-h_{+}} +
(h_{+}\leftrightarrow h_{-}),         \label{lim2}
\end{equation}
where $c_{+}$ is a constant and $\cos \Theta$ is given by (\ref{ang})
in four dimensions.

As in \cite{Strom1,Klem1} we also impose the following boundary
condition for $\Phi$ at the boundary $I^{-}$
\begin{equation}
\lim_{r\rightarrow \infty}\Phi(t,r,\theta,\phi) =
r^{-h_{-}}\Phi_{-}(t,\theta,\phi).                       \label{lim3}
\end{equation}

However, as stressed by Strominger \cite{Strom1} the boundary
condition (\ref{lim3}) is not mandatory, and we may also choose 
$\Phi(t,r,\theta,\phi)=r^{-h_{+}}\Phi_{+}(t,\theta,\phi)$ at infinity.
This leads to similar correlators, obtained by the change $h_{+}
\leftrightarrow h_{-}$. Thus, from this point onwards we simply use
$h$, and at the end we shall obtain the full set of solutions chosing
either value of $h$.

The use of (\ref{lim2}) and (\ref{lim3}) in (\ref{lim1}) leads to
\begin{eqnarray}
D(\Phi,\Phi^{\prime})&\equiv&\lim_{r\rightarrow \infty}\int dt d\theta
d\phi dt^{\prime}d\theta^{\prime}d\phi^{\prime}\sin\theta\sin
\theta^{\prime}\frac{(rr^{\prime})^{2}}{a^{2}}\times \nonumber \\
&& \left[\Phi(x)\flex\partial_{r_{\ast}}G(x,x^{\prime})\flex\partial_{r^{\prime}_{\ast}}\Phi(x^{\prime})\right]_{r=r^{\prime}}\,=
\label{int}
\end{eqnarray}
\begin{equation}
= \int dt d\theta d\phi dt^{\prime} d\theta^{\prime} d\phi^{\prime}
\frac{\sin\theta\sin\theta^{\prime}\Phi_{-}(t,\theta,\phi)\Phi_{-}(t^{\prime},\theta^{\prime},\phi^{\prime})}
{[\cosh(\frac{t-t^{\prime}}{a}) -(\sin \theta \sin\theta^{\prime}
\cos (\phi-\phi^{\prime}) + \cos \theta \cos \theta^{\prime})]^{h}}.
\label{int1}
\end{equation}

Now we substitute above the {\it Ansatz} $\Phi_{-}(t,\theta,\phi)=
e^{i\omega t}Y_{\ell}^{m}(\theta,\phi)$. We note that up to this point
there is {\it a priori} no relationship here between the parameter
$\omega$ in this {\it Ansatz} and that denoting the quasi-normal modes
from the bulk perturbations. It follows that

\begin{equation}
D_{\omega\omega^{\prime}} =\int dt d\theta d\phi dt^{\prime} d\theta^{\prime}d\phi^{\prime}
\frac{\sin\theta\sin\theta^{\prime}  
Y^{m}_{\ell}(\theta,\phi)Y^{m^{\prime}}_{\ell^{\prime}}(\theta^{\prime},
\phi^{\prime})e^{i\omega t}e^{i\omega^{\prime}t^{\prime}}}{[\cosh(\frac{t-t^
{\prime}}{a})-(\sin \theta \sin \theta^{\prime}\cos (\phi-\phi^{\prime}) + 
\cos \theta \cos \theta^{\prime})]^{h}},                     \label{int2}
\end{equation}
which can be suitably written in the form (up to a constant factor, $C$,
coming from the normalization of the spherical harmonics) 
\begin{eqnarray}
D_{\omega\omega^{\prime}}&=&C\delta_{mm^{\prime}}\delta(\omega
-\omega^{\prime})\int d\tau d\theta
d\theta^{\prime}d\varphi\sin\theta\sin\theta^{\prime}
P^{m}_{\ell}(\cos\theta)P^{m}_{\ell^{\prime}}(\cos\theta^{\prime})\times
\nonumber \\
&& e^{im\varphi}e^{i\omega\tau}\left[\cosh(\frac{\tau}{a}) -(\sin\theta\sin\theta^{\prime}\cos\varphi+\cos\theta\cos\theta^{\prime})\right]^{-h}\quad,
\label{int3}
\end{eqnarray}
where $\varphi=\phi-\phi^{\prime}$ and $\tau=t-t^{\prime}$.

It is our aim here to display the spectrum of perturbations, that
is, the poles of (\ref{int3}) in $\omega$. This is not too
difficult, since each pole is caracterized by the asymptotic behaviour
of the integrand for $\tau \rightarrow\infty$. Indeed, a pole $1/(\omega^{2}-\omega_{0}^{2})$ corresponds to
an asymptotic behaviour $\exp(i\omega_{0}\tau)$. Thus, by expanding 
(\ref{int3}) in powers of
\begin{equation}
\epsilon=\frac{1}{\cosh(\frac{\tau}{a})}
(\sin \theta \sin \theta^{\prime}\cos \varphi + \cos \theta \cos
\theta^{\prime})\quad,                 \label{potx}
\end{equation}
we have
\begin{eqnarray}
&&D_{\omega\omega^{\prime}}= C\delta_{mm^{\prime}}\delta(\omega - \omega^{\prime})\int
\frac{d\tau e^{i\omega\tau}}{\cosh(\frac{\tau}{a})^{h}}
\left [a_{0} + a_{1}\frac{h}{\cosh(\frac{\tau}{a})} + 
a_{2}\frac{h(h+1)}{2!\cosh^{2}(\frac{\tau}{a})} + \right.
\nonumber \\
&& \left. +\,a_{3}\frac{h(h+1)(h+2)}{3!\cosh^{3}(\frac{\tau}{a})} +
...\right]\,,
\label{expssao}
\end{eqnarray}
where
\begin{eqnarray}
a_{0}=\int d\theta d\theta^{\prime}\sin\theta\sin\theta^{\prime}
P^{m}_{\ell}(\cos\theta)P^{m}_{\ell^{\prime}}(\cos \theta^{\prime})
\int d\varphi\,e^{im\varphi} \,,
\end{eqnarray}
\begin{eqnarray}
a_{1}&=&\int d\theta d\theta^{\prime}\sin\theta\sin\theta^{\prime}
P^{m}_{\ell}(\cos\theta)P^{m}_{\ell^{\prime}}(\cos \theta^{\prime})
\int d\varphi e^{im\varphi}\times \nonumber \\
&& \left[
\sin\theta\sin\theta^{\prime}\cos\varphi + \cos\theta\cos
\theta^{\prime}\right]\,;\,... 
\end{eqnarray}
\begin{eqnarray}
a_{N}&=&\int d\theta d\theta^{\prime}\sin\theta\sin\theta^{\prime}
P^{m}_{\ell}(\cos\theta)P^{m}_{\ell^{\prime}}(\cos \theta^{\prime})
\int d\varphi e^{im\varphi}\times \nonumber \\
&&\left[\sin\theta\sin\theta^{\prime}\cos\varphi + \cos\theta\cos
\theta^{\prime}\right]^{N}\,...\,.
\end{eqnarray}

It is simple to solve the integrals in $\varphi$ by
means of the orthogonality relation of
$exp(im\varphi)$. This will fix the value of
$m$ at each term in the expansion. The integrals in $\theta$ and
$\theta^{\prime}$ are a little more involved, but they can be handled
by using standard integrals (as, {\it e.g.}, in \cite{gradst})
involving the $P^{m}_{\ell}(\theta)\,$'s themselves and contributions of
sines and cossines. For example, up to the second-order term in the
$\epsilon$-expasion we find integrals of the type
\begin{eqnarray}
a_{0}&=& 2\pi\delta_{m,0}\int_{-1}^{1}dx^{\prime}P^{0}_{\ell^{\prime}}
(x^{\prime})\int_{-1}^{1}dxP^{0}_{\ell}(x)\,,
\\
a_{1}&=&\pi\delta_{m,1}\int_{-1}^{1}dx^{\prime}P^{1}_{\ell^{\prime}}(x^{\prime})
(1-x^{\prime 2})^{1/2}\int_{-1}^{1}dxP^{1}_{\ell}(x)(1-x^{2})^{1/2} + \nonumber \\ 
&&2\pi\delta_{m,0}\int_{-1}^{1}dx^{\prime}P^{0}_{\ell^{\prime}}
(x^{\prime})x^{\prime}\int_{-1}^{1}dxP^{0}_{\ell}(x)x\,,
\\
a_{2}&=&
\frac{\pi}{2}\delta_{m,2}\int_{-1}^{1}dx^{\prime}P^{2}_{\ell^{\prime}}(x^{\prime})(1-x^{\prime2})\int_{-1}^{1}dxP^{2}_{\ell}(x)(1-x^{2}) 
+ \nonumber \\
&&2\pi\delta_{m,1}\int_{-1}^{1}dx^{\prime}P^{1}_{\ell^{\prime}}(x^{\prime})x^{\prime}(1-x^{\prime
2})^{1/2}\int_{-1}^{1}dxP^{1}_{\ell}(x)x(1-x^{2})^{1/2} + \nonumber \\
&&2\pi\delta_{m,0}\int_{-1}^{1}dx^{\prime}P^{0}_{\ell^{\prime}}
(x^{\prime})x^{\prime 2}\int_{-1}^{1}dxP^{0}_{\ell}(x)x^{2} +
\nonumber \\
&&\pi\delta_{m,0}\int_{-1}^{1}dx^{\prime}P^{0}_{\ell^{\prime}}(x^{\prime})(1-x^{\prime2})\int_{-1}^{1}dxP^{0}_{\ell}(x)(1-x^{2})+\nonumber \\
&&\frac{\pi}{2}\delta_{m,-2}\int_{-1}^{1}dx^{\prime}P^{-2}_{\ell^{\prime}}(x^{\prime})(1-x^{\prime2})\int_{-1}^{1}dxP^{-2}_{\ell}(x)(1-x^{2})\,, 
\end{eqnarray}
where $x=\cos\theta$ and $x=\cos\theta^{\prime}\,$.

The computation of the above integrals fixes the values of $\ell$ and
$\ell^{\prime}$ that leads to a non-vanishing contribution for the
coefficients $a_{0}, a_{1}\,$, and $a_{2}$\,. This can be easily done
using the orthogonality of the $P^{m}_{\ell}(\theta)\,$'s, for some
integrals, and resorting to a table for others. Almost all the
integrals appearing in $a_{0}, a_{1}\,$, and $a_{2}$ can be expressed as
a product of $P^{m}_{\ell}\,$'s with the same index $m$. This allow us
to obtain for those integrals
\begin{eqnarray}
\int_{-1}^{1}dxP^{0}_{\ell}(x)&=&
\frac{2}{2\ell+1}\,\delta_{\ell,\,0}\quad,
\\
\int_{-1}^{1}dxP^{1}_{\ell}(x)(1-x^{2})^{1/2}&=&
\frac{2(\ell+1)!}{(2\ell+1)(\ell-1)!}\,\delta_{\ell,\,1}\quad,
\\
\int_{-1}^{1}dxP^{0}_{\ell}(x)x &=&
\frac{2}{2\ell+1}\,\delta_{\ell,\,1}\quad, 
\\
\int_{-1}^{1}dxP^{2}_{\ell}(x)(1-x^{2})&=&
\frac{2(\ell+2)!}{3(2\ell+1)(\ell-2)!}\,\delta_{\ell,\,2}\quad,
\\
\int_{-1}^{1}dxP^{1}_{\ell}(x)x(1-x^{2})^{1/2}&=&
\frac{2(\ell+1)!}{3(2\ell+1)(\ell-1)!}\,\delta_{\ell,2}\quad
\\
\int_{-1}^{1}dxP^{0}_{\ell}(x)x^{2}&=& 
\frac{4}{3(2\ell+1)}\delta_{\ell,2} +
\frac{4}{3(2\ell+1)}\,\delta_{\ell,\,0}\quad,
\end{eqnarray}
from where we see that only $\ell, \ell^{\prime}= 0;1;1;2;2;(0;2)$
contribute for the above integrals, respectively. The integrals containing
$P^{-m}_{\ell}$ can be solved by means of the relation
$P^{-m}_{\ell}=(-1)^{m}\frac{(\ell-m)!}{(\ell+m)!}P^{m}_{\ell}$.

It remains the integral $\int_{-1}^{1}dxP^{0}_{\ell}(x)(1-x^{2})\,$, that
appears in $a_{2}$, which cannot be written as a product of the
$P^{m}_{\ell}\,$'s with the same $m\,$. In this case, we use
\cite{gradst} to obtain
\begin{eqnarray} 
&&\int_{-1}^{1}dx^{\prime}P^{0}_{\ell^{\prime}}(x^{\prime})(1-x^{\prime 2})
\int_{-1}^{1}dxP^{0}_{\ell}(x)(1-x^{2})= \nonumber \\
&&\frac{\pi}{\Gamma(2+\frac{\ell^{\prime}}{2}+\frac{1}{2})\Gamma(2-\frac
{\ell^{\prime}}{2})\Gamma(\frac{\ell^{\prime}}{2}+1)\Gamma(-\frac{\ell^
{\prime}}{2}+\frac{1}{2})} \times (\ell^{\prime}\leftrightarrow \ell)\,.       \label{intgams}
\end{eqnarray}

We conclude that only $\ell, \ell^{\prime}= 0,2$ contribute with
a non-vanishing value for the integrals in (\ref{intgams}).

It is easy to verify that the non-vanishing contributions for the
integrals in each higher-order term in the expansion (\ref{expssao}) 
are similar to
those shown above. In fact, for the third-order term,  we
have the contributions $\ell, \ell^{\prime}=1,3\,$, and as the
inspection of a generic term $a_{N}$ in the expansion shows, the
contributions of $\ell, \ell^{\prime}$ will be of the type 
$\ell, \ell^{\prime}= N, N-2, N-4,...\,$. This allows us to
write (\ref{expssao}) as
\begin{equation}
D_{\omega\omega^{\prime}}=\delta(\omega - \omega^{\prime})
\sum_{N=0}c_{N}\int d\tau e^{i\omega \tau}\left[\cosh
\frac{\tau}{a}\right]^{-(h+l+2N)}\,.                    \label{soma1} 
\end{equation}

This integral has poles at 
\begin{equation}
\omega=\pm\frac{i}{a}(h_{\pm} + l + 2n)\quad,                \label{Spect}
\end{equation}
in view of the behaviour of the integrand for large values of
$\tau\,$.

The spectrum (\ref{Spect}) coincides exactly with the quasi-normal
frequencies (\ref{eqome2}) obtained in perturbations of the bulk of the 
de Sitter space-time, except only for the values $\pm \xi \pm 1/2\,$,
where $\xi= \sqrt{9-4\mu^{2}a^{2}}\,$. This can be easily verified by
inspection of the bulk spectrum, given by (\ref{eqome2}-\ref{eqome3}).

\section{Concluding remarks}
$ $

We have shown that the quasi-normal modes arising from a scalar perturbation
of the de Sitter space are, with exception of only four of them, contained
in the spectrum of two-point function of the corresponding
three-dimensional conformal field theory at the boundary.

Although the computation of the two-point correlator has been
performed in four-dimensional 
de Sitter space, the results can be generalized to
$D$ dimensions, where presumably the quasi-normal modes obtained in
bulk de
Sitter space are, with the exception of a small number of them,
contained in the spectrum of the corresponding $(D-1)$-dimensional CFT
at the boundary. In fact, this seems to occur, since the form of the
two-point correlator as given by (\ref{int2}) can be directly
generalized to $D$ dimensions. In this case, we will have to handle
with hyperspherical harmonics and the denominator of (\ref{int2})
takes the form $[\cosh\frac{\tau}{a}-\cos\Theta]^{-h}\,$, where 
$\Theta=\Theta(\Omega,\Omega^{\prime})$ is the geodesic distance
between two points on the unit sphere $S^{D-2}\,$.

Our results give directions to build foundations of an extension
of the celebrated  AdS/CFT correspondence to the de Sitter space. We
do not know the interpretation of the small number of states left out
of the CFT spectrum, but for extensive magnitudes, such as entropy, they
presumably do not matter.

At last, we stress the fact, already well signalized, for example, by 
Strominger 
\cite{Strom1}, that the whole de Sitter space-time cannot be probed by a
single observer, and describing the whole de Sitter space-time
corresponds to describing both sides of a black hole's event
horizon. In spite of the discussion involving a region smaller than the
full de Sitter space, we have shown a striking evidence that a CFT
describes well the holographic projection of the bulk space, thus
providing strong support for a dS/CFT correspondence, since bulk
eigenmodes are fully described in the region of space probed by a
single observer.


\bigskip
\noindent {\bf Acknowledgements}\par
The authors would like to thank {\bf CNPq-Brazil} ({\it Conselho
Nacional de Desenvolvimento Cient\'\i fico e Tecnol\'ogico}) and
{\bf FAPESP-Brazil} ({\it Funda\c{c}\~ao de Amparo \`a Pesquisa
no Estado de S\~ao Paulo}) for financial support.


\begin{thebibliography}{99}
\bibitem{lbpos}
B. P. Schmidt {\it et al.}, Astrophys. J. {\bf 507}, 46 (1998);\\
A. G. Riess {\it et al.}, Astron. J. {\bf 116}, 1009 (1998);\\  
S. Perlmutter {\it et al.}, Astrophys. J. {\bf 517}, 565 (1999);\\
S. Perlmutter, in {\it Proc. of the 19th Intl. Symp. on Photon and
Lepton Interactions at High Energy LP99}, ed. J.A. Jaros and
M. E. Peskin, Int. J. Mod. Phys. A {\bf 15S1}, 715 (2000).
\bibitem{cobe}
P. de Bernardis {\it et al.}, Nature {\bf 404}, 955 (2000);\\
R. Stompor {\it et al.}, Astrophys. J. {\bf 561}, L7 (2001).  
\bibitem{bousso1}
R. Bousso, hep-th/0205177.
\bibitem{malda}
J. Maldacena, Adv. Theor. Math. Phys. {\bf 2}, 231 (1998);\\
S. S. Gubser, I. Klebanov, and A. Polyakov, Phys. Lett. B {\bf 428},
105 (1998); \\
E. Witten,  Adv. Theor. Math. Phys. {\bf 2}, 253 (1998).
\bibitem{severalads}
O. Aharony, S. S. Gubser, J. Maldacena, H. Ooguri,
and Y. Oz, Phys. Rep. {\bf 323}, 183 (2000).
\bibitem{witten}E. Witten, hep-th/0106109.
\bibitem{Strom1}
A. Strominger, J.H.E.P {\bf 0110}, 034 (2001), hep-th/0106113.
\bibitem{severalds}
S. Nojiri and S. D. Odintsov, Phys. Lett. B {\bf 531}, 143 (2002); 
S. Nojiri, S. D. Odintsov, and S. Oguchi, Phys. Rev. D {\bf 66}, 023522 (2002).\\
H. W. Lee and Y. S. Myung, Phys. Lett. B {\bf 537}, 117 (2002);\\
B. McInnes, Nucl. Phys. B {\bf 627}, 311 (2002);\\
M. Cvetic, S. Nojiri, and S. D. Odintsov, Nucl. Phys. B {\bf 628}, 295
(2002);\\
Z. Chang and C.-B. Guan, hep-th/0204014;\\
B. C. da Cunha, hep-th/0208018.
\bibitem{GibHawk}
G. W. Gibbons and S. W. Hawking, Phys. Rev. D {\bf 15}, 2738 (1977). 
\bibitem{ablimbin}
E. Abdalla, B. Wang, A. Lima-Santos, and W. G. Qiu, Phys. Lett. B {\bf
538}, 435 (2002).
\bibitem{brady}
P. Brady, C. M. Chambers, W. Krivan, and P. Laguna, Phys. Rev. D 
{\bf 55}, 7538 (1997).
\bibitem{abdmolsaa}
E. Abdalla, C. Molina, and A. Saa ({\it in preparation}).
\bibitem{ref5deablimbin}
J. S. F. Chan and R. B. Mann, Phys. Rev. D{\bf 59}, 064025 (1999);\\
J. S. F. Chan and R. B. Mann, Phys. Rev. D{\bf 55}, 7546 (1997).
\bibitem{ref6deablimbin}
G. T. Horowitz and V. E. Hubeny, Phys. Rev. D{\bf 62}, 024027 (2000);\\
G. T. Horowitz, Class. Quant. Grav. {\bf 17}, 1107 (2000).
\bibitem{ref7-8deablimbin}
B. Wang, C. Y. Lin, and E. Abdalla, Phys. Lett.B{\bf 481}, 79 (2000);\\
B. Wang, C. Molina, and E. Abdalla, Phys. Rev. D{\bf 63}, 084001 (2001);\\
J. M. Zhu, B. Wang, and E. Abdalla, Phys. Rev. D{\bf 63}, 124004 (2001).
\bibitem{ref9deablimbin}
V. Cardoso and J. P. S. Lemos, Phys. Rev. D{\bf 63}, 124015 (2001).
\bibitem{ref10deablimbin}  
B. Wang, E. Abdalla, and R. B. Mann, Phys. Rev. D{\bf 65}, 084006 (2002).
\bibitem{btz}
M. Ba\~{n}ados, C. Teitelboim, and J. Zanelli, Phys. Rev. Lett. {\bf
69}, 1849 (1992).
\bibitem{birmg}
D. Birmingham, I. Sachs, and S. N. Solodukhin, Phys. Rev. Lett. {\bf
88}, 
151301 (2002). 
\bibitem{Leshouches}
M. Spradlin, A. Strominger, and A. Volovich, {\it Les Houches Lectures
on de Sitter Space}, hep-th/0110007.
\bibitem{Klem1}
D. Klemm, Nucl. Phys. B {\bf 625}, 295 (2002).
\bibitem{bateman}
{\it  Higher Transcedental Functions, V. {\bf II} (Bateman Manuscript
Project)}, ed. by A. Erdélyi (McGraw-Hill, New York, 1953).
\bibitem{gradst}
I. S. Gradshteyn and I. M. Ryzhik, {\it Tables of Integrals, Series
and Products}, Academic Press (2000).
\bibitem{Can}
P. Candelas and D. J. Raine, Phys. Rev. D {\bf 12}, 965 (1975).
\end{thebibliography}
\end{document}